\documentclass[%
 rsi,
 amsmath,amssymb,
 reprint,%
 superscriptaddress,
]{revtex4-1}

\usepackage{enumitem}
\usepackage{graphicx}
\usepackage{dcolumn}
\usepackage{bm}

\usepackage[utf8]{inputenc}
\usepackage[T1]{fontenc}
\usepackage{mathptmx}
\usepackage{amsmath}
\usepackage{etoolbox}
\usepackage{comment}
\usepackage[dvipsnames]{xcolor}
\usepackage{ amssymb }
\usepackage[caption=false]{subfig}
\usepackage[title]{appendix}
\usepackage{enumerate}
\usepackage{physics}
\makeatletter
\def\@email#1#2{%
 \endgroup
 \patchcmd{\titleblock@produce}
  {\frontmatter@RRAPformat}
  {\frontmatter@RRAPformat{\produce@RRAP{*#1\href{mailto:#2}{#2}}}\frontmatter@RRAPformat}
  {}{}
}%

\makeatother
\begin{document}
\preprint{AIP/123-QED}

\title{Experimental advances with the QICK (Quantum Instrumentation Control Kit) \\ for superconducting quantum hardware}


\author{Chunyang Ding}
\affiliation{Department of Physics and Applied Physics, Stanford University, Stanford CA, 94305}
 \author{Martin Di Federico}
 \affiliation{Fermi National Accelerator Laboratory, Batavia IL, 60510}
\author{Michael Hatridge}
 \affiliation{Department of Physics \& Astronomy, University of Pittsburgh, Pittsburgh PA, 15213}
\author{Andrew Houck}
 \affiliation{Department of Electrical Engineering, Princeton University, Princeton NJ, 08544}
  \author{Sebastien Leger}
\affiliation{Department of Physics and Applied Physics, Stanford University, Stanford CA, 94305}
 \author{Jeronimo Martinez}
 \affiliation{Department of Electrical Engineering, Princeton University, Princeton NJ, 08544}
 \author{Connie Miao}
\affiliation{Department of Physics and Applied Physics, Stanford University, Stanford CA, 94305}
 \author{David I Schuster}
\affiliation{Department of Physics and Applied Physics, Stanford University, Stanford CA, 94305}
\author{Leandro Stefanazzi}
 \affiliation{Fermi National Accelerator Laboratory, Batavia IL, 60510}
  \author{Chris Stoughton}
 \affiliation{Fermi National Accelerator Laboratory, Batavia IL, 60510}
 \author{Sara Sussman}
 \affiliation{Fermi National Accelerator Laboratory, Batavia IL, 60510}
 \author{Ken Treptow}
 \affiliation{Fermi National Accelerator Laboratory, Batavia IL, 60510}
\author{Sho Uemura}
 \affiliation{Fermi National Accelerator Laboratory, Batavia IL, 60510}
 \author{Neal Wilcer}
 \affiliation{Fermi National Accelerator Laboratory, Batavia IL, 60510}
\author{Helin Zhang}
\affiliation{Research Laboratory of Electronics, Massachusetts Institute of Technology, Cambridge, MA 02139, USA}

 \author{Chao Zhou}
 \affiliation{Department of Physics \& Astronomy, University of Pittsburgh, Pittsburgh PA, 15213}

\author{Gustavo Cancelo*}
 \affiliation{Fermi National Accelerator Laboratory, Batavia IL, 60510}

 \email{cancelo@fnal.gov.}
\date{\today}

\begin{abstract}
The QICK is a standalone open-source qubit controller that was first introduced in 2022. In this follow-up work, we present recent experimental use cases that the QICK uniquely enabled for superconducting qubit systems. These include multiplexed signal generation and readout, mixer-free readout, pre-distorted fast flux pulses, and phase-coherent pulses for parametric operations, including high-fidelity parametric entangling gates. We explain in detail how the QICK was used to enable these experiments.
\end{abstract}

\maketitle


\section{Introduction}\label{Introduction}

Achieving high-fidelity qubit control and readout requires significant RF engineering and digital signal processing. To control superconducting qubits, for example, signals range from DC to upwards of 10 GHz, with complex envelopes. Signals must jump in frequency while maintaining phase coherence and meeting very precise timing requirements. The total control pulse sequence must be significantly faster than the coherence time of any qubit in the system (typically microseconds), and the latency between any two pulses must be controllable (zero latency is often required). In addition, low-latency conditional logic is required for feedback and feedforward. Each qubit typically needs several dedicated lines for control and readout, and hundreds of qubits must be operated in parallel to perform a nontrivial quantum computation ~\cite{Bravyi2022,Acharya2023}.

The Quantum Instrumentation Control Kit (QICK) is an open-source system which uses the AMD-Xilinx RFSoC to deliver these functions.
The RFSoC ~\cite{rfsocwebsite, Crockett2023} is a system-on-chip that incorporates high-speed DACs and ADCs and has been adopted for qubit control by both the research community~\cite{Gebauer2021,park2021icarusq} and commercial industry~\cite{imp}.
RFSoC-based open-source qubit controllers~\cite{Xu2023, Stefanazzi2022} are a growing alternative to commercial solutions~\cite{qumachines,zurichinst}.


The first-generation Quantum Instrumentation Control Kit (QICK) was introduced in Ref.~\cite{Stefanazzi2022} as an open-source superconducting qubit controller running on an RFSoC evaluation board. The QICK firmware is a scaffold with a modular timed-processor and signal generation/readout blocks that the user can combine in many configurations. The QICK firmware and software can be downloaded from a public GitHub repository~\cite{QICKrepo, QICKdocs}. The original RFSoC chip supported by QICK was the Gen 1 ZU28DR running on the AMD-Xilinx ZCU111 evaluation board~\cite{zcu111}. In this paper, we describe QICK running on the Gen 3 ZU49DR RFSoC chip and its evaluation boards~\cite{zcu216,PYNQ4x2}.

\begin{figure}[hb!]
\begin{center}
\includegraphics[width= \columnwidth]{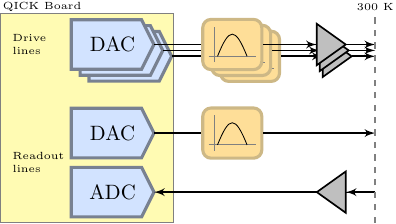}
        \caption{\textbf{A typical superconducting qubit control loop using the QICK on the ZCU216 board.} Drive pulses are filtered then amplified before being sent to the fridge. The readout pulse is filtered before being sent to the fridge. After the fridge, the readout pulse is amplified before being directly read into the ADC.}\label{fig:ZCU216_ControlLoop}
\end{center}
\end{figure}

The Gen 3 ZCU216 evaluation board is of particular interest because of its 16 DACs running at 9.85~GS/s.
The large number of high-speed DACs makes the ZCU216 useful for controlling multi-qubit systems. In addition, the ZCU216 also has 16 ADCs running at 2.5~GS/s. For this next-generation hardware, QICK firmware was developed that enables direct, mixer-free generation of pulses for qubit drives and multiplexed readout up to 10 GHz. Additionally, the 100~ps time resolution enables precisely defined qubit control pulses. Phase coherence is maintained across all ZCU216 channels, reducing the need for external triggers and synchronized clocks across different equipment.


Fig.~\ref{fig:ZCU216_ControlLoop} shows a typical control loop for measuring a superconducting qubit system using the QICK running on the ZCU216 evaluation board. This scheme eliminates the warm analog components used in superconducting qubit control besides amplification and filtering. A QICK RF companion board for the ZCU216 has been designed and is currently in production. Due to the large analog bandwidth of the ZCU216, its QICK RF companion board is correspondingly simpler than that of the ZCU111~\cite{Stefanazzi2022}.
As a cost-effective or educational alternative, QICK can run on the RFSoC4x2 evaluation board which is sold with special academic pricing. The RFSoC4x2 has 2 DACs running at 9.85~GS/s and 4 ADCs running at 5~GS/s.

The new capabilities we describe in this paper were also facilitated by an upgraded Python software library~\cite{QICKrepo}. The QICK software is based on the open-source Xilinx PYNQ (Python productivity on Zynq) operating system~\cite{pynqwebsite,Crockett2023}, which is distributed as disk images from AMD-Xilinx~\cite{pynqboards} and the QICK team~\cite{ZCU216-PYNQ}.
Every QICK firmware block has a corresponding PYNQ driver that abstracts the hardware details. These drivers are automatically attached to the firmware block without any user intervention and are maintained as part of the QICK library. On top of the driver abstraction, the QICK library has a powerful parser and automatic connectivity detection algorithm that creates data structures to efficiently handle all the available resources, such as different types of signal generator and readout blocks that all run at different speeds.

 Several experiments have been already published using the QICK to control superconducting quantum hardware~\cite{Bryon2023,Martinez2023,Xie2023,Anferov2023,CruzMartinez2023, Efthymiou2023, Zhang2023}. In this paper we present recent experimental use cases that were uniquely enabled by the QICK. We highlight some key capabilities (Section~\ref{sec:capabilities}) and relevant examples: multiplexed signal generation and readout (Section~\ref{sec:expt_qram}), pre-distorted fast flux pulses (Section~\ref{sec:expt_pre-distorted}), phase-coherent parametric control (Section~\ref{sec:expt_ParametricMeasurement}), and phase-sensitive parametric entangling gates (Section~\ref{sec:expt_ParametricEntangler}). We conclude in Section~\ref{Conclusion} with remarks on further improvements planned for the QICK.

\section{Signal generation and readout}\label{sec:capabilities}


QICK has an extensive library of ``signal generator'' and ``readout'' firmware blocks which connect to the DACs and ADCs of the FPGA. 
These blocks are compatible with any of the supported RFSoC boards and any DAC or ADC sampling rate supported by the FPGA, and can be combined as needed for the requirements of different experiments. In addition, we have released compiled and tested versions of the firmware, enabling researchers to use high-level Python code to immediately control superconducting qubit systems.  

\subsection{Multiplexed signal generation and readout}\label{sec:multiplexed}

\begin{figure}[htbp!]
        \begin{center}
        \includegraphics[width=\columnwidth]{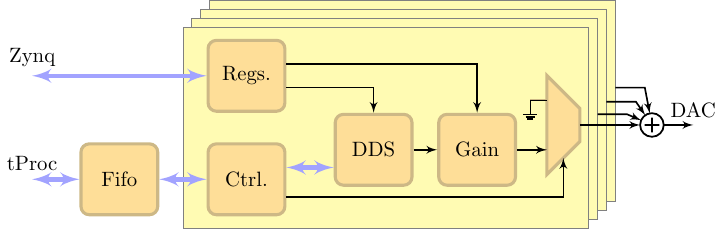}
        \caption{\textbf{Multiplexed signal generation.} Frequency multiplexing is achieved by adding two or more single-frequency signal generators before the DAC module.} \label{fig:muxSG}
\end{center}
\end{figure}

Multiple signals can be frequency-multiplexed on a single DAC output.
Fig.~\ref{fig:muxSG} shows one approach (used in Sec.~\ref{sec:expt_qram}), where independent DDS channels are digitally summed before the DAC.
For multiplexing large numbers of tones, a different approach using polyphase filter bank (PFB)~\cite{harris2022multirate} is possible.

%

\begin{figure}[!hb]
	\begin{center}
\includegraphics[width=\columnwidth]{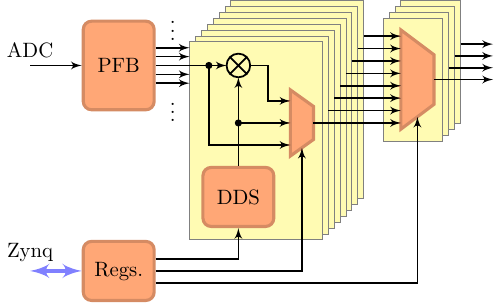}
	\end{center}
	\vspace{-0.6cm}
        \caption{\textbf{Multiplexed readout.} A polyphase filter bank (PFB) digitally demultiplexes the ADC samples into 8 channels with 50\% overlap, to avoid gain losses over the entire bandwidth.} \label{fig:Readout}
\end{figure}

Figure~\ref{fig:Readout} shows a multiplexed readout, also used in Sec.~\ref{sec:expt_qram}.
The ADC data stream has a bandwidth equal to the Nyquist frequency; a polyphase filter bank (PFB) applies an array of bandpass filters to split this spectrum evenly into channels.
Each channel has its own DDS oscillator which can be used to demodulate a signal that falls in its frequency range.
In the current implementation with 8 channels and 4 outputs, each channel has a width of 1/16 the sampling frequency and up to 4 channels can be read out simultaneously.

\subsection{Full-speed and interpolated envelopes}\label{sec:envelopes}

Waveform envelopes, such as Gaussian, DRAG, triangular, or user-provided, represent general waveform shapes which the signal generator can use to parametrically create pulses with arbitrary frequency, amplitude, and phase. 

\begin{figure}[!ht]
\begin{center}
\includegraphics[width=\columnwidth]{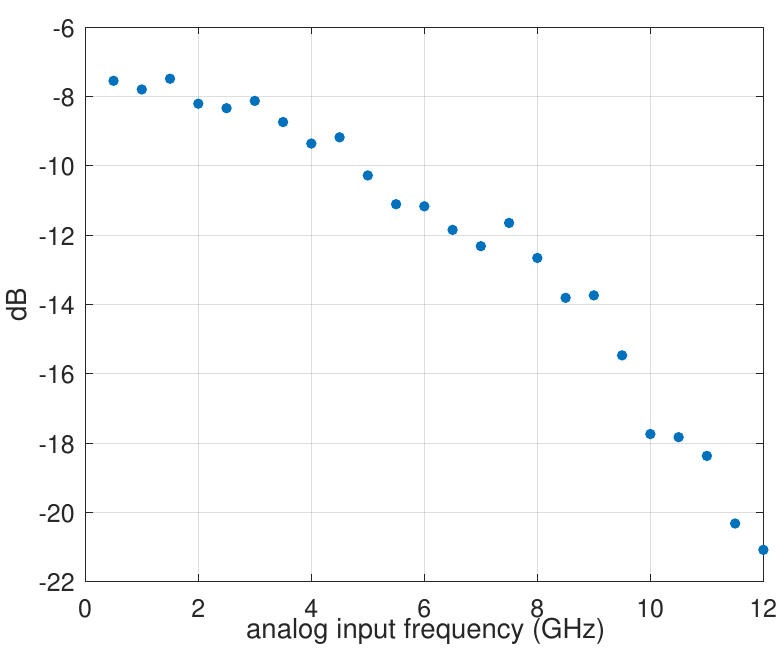}\\
\includegraphics[width=\columnwidth]{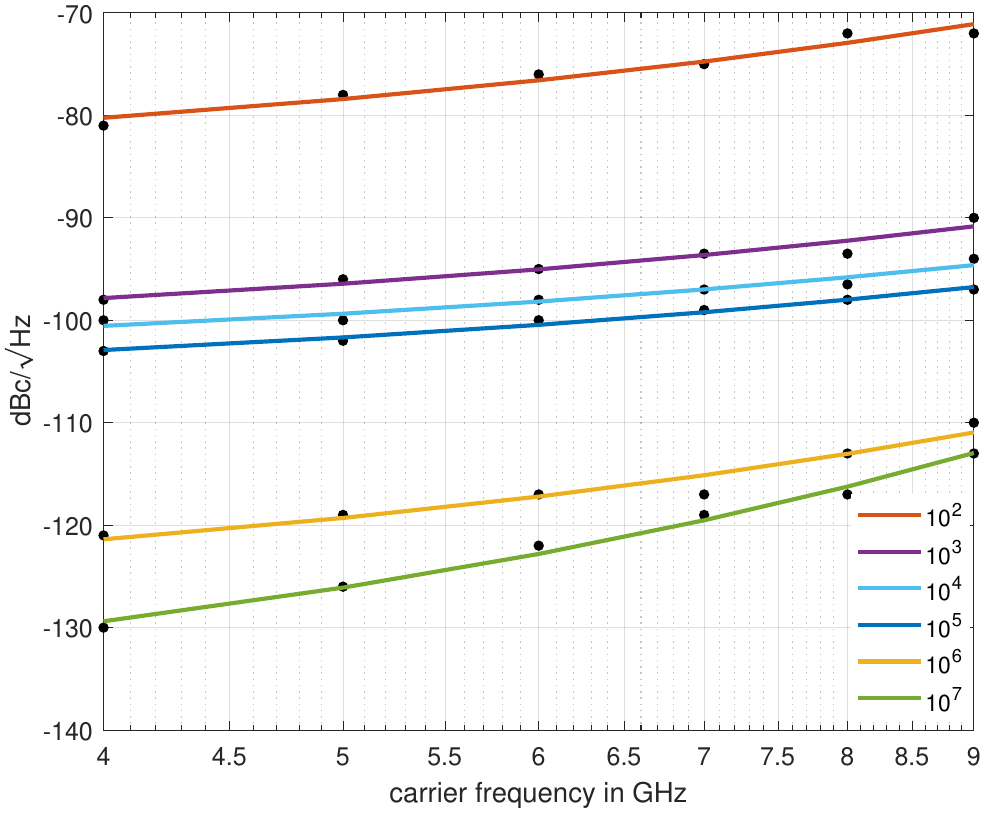}
\caption{(top) ZCU216 ADC analog gain as a function of analog input frequency. The transfer function is measured in 0.5 GHz steps at a constant input power of -7 dBm. The ADC sample rate is 2.4572 GHz. (bottom) ZCU216 ADC phase noise measured at $-1dBFS$ of power, in the range from 4 to 9 GHz $f_c$ and at delta frequencies from 100 Hz to 10 MHz from $f_c$.}
\label{fig:ADC_gain_noise} 
\end{center}    
\end{figure}

In the standard ``full-speed'' signal generator, the envelope sample rate is equal to the DAC sample rate.
This enables precise control of the pulse shape and sub-nanosecond time resolution, as used in Sec.~\ref{sec:expt_pre-distorted}.
Oscillating envelopes can be used for fast chirps or detuned pulses which can use the full DAC bandwidth.

We have also developed an interpolated signal generator where the envelope rate is 1/16 the DAC rate, and time-domain interpolation is used to upsample the envelope to the DAC rate~\cite{DSP_book}.
This uses both the envelope memory and the FPGA logic more efficiently, and is appropriate for the typical case of near-DC envelopes with narrow bandwidth.


\subsection{Phase coherence}\label{sec:phase}

All generators and readouts use digital mixers, where a complex carrier oscillator is produced by direct digital synthesis (DDS) and is multiplied by an envelope (for a signal generator) or the ADC data stream (for a readout).
Because the DDS oscillators are purely numerical and driven by a common FPGA clock, their relative phases are completely predictable.
This scheme allows QICK signal generators to preserve phase coherence across multiple pulses of the same frequency while frequency-hopping~\cite{Stefanazzi2022}.
We have now added the capability (used in Sec.~\ref{sec:expt_ParametricMeasurement}) to apply a synchronous phase reset to multiple signal generators and/or readouts, which can be used to set a fixed relative phase between pulses of differing frequencies.

\subsection{Mixer-free readout}\label{MixerFree}

The ZCU216 DACs have a maximum sampling frequency of 9.85~GS/s, and the ADCs have a maximum sampling frequency of 2.5~GS/s.
However, this does not impose a hard limit on the frequencies of signals that can be used; both the DACs and ADCs are limited only by their analog bandwidths.
The RFSoC DACs can be configured for normal mode (zero-order hold, optimal for the first Nyquist zone) or ``mix-mode'' (boosted power in the second and third Nyquist zones).


As shown in Fig.~\ref{fig:ZCU216_ControlLoop}, the QICK readout input uses direct sampling to simplify the analog electronics and avoid the use of analog mixers~\cite{proakis,DSP_book}.
An analog bandpass filter should be applied to the input signal to remove noise from other Nyquist zones, and a preamplifier should be used to compensate for the reduced ADC gain at higher frequencies (shown in Fig.~\ref{fig:ADC_gain_noise}(a)).
Fig.~\ref{fig:ADC_gain_noise}(b) shows that this prescription performs well: the noise near the signal frequency is dominated by phase noise, and further away is set by the white noise floor.
In the worst case measured (100 Hz separation from a 9 GHz signal), the phase noise is still below -70~dBc, which is low enough for typical quantum measurements.


\section{Four qubit simultaneous readout using MUX generation and readout}\label{sec:expt_qram}

The multiplexed signal generator and readout described in Sec.~\ref{sec:multiplexed} makes experiments more hardware-efficient and less prone to calibration and user error. To illustrate this, we introduce an experiment realized in the Schuster Lab at Stanford.

\begin{figure}[!htbp]
	\begin{center}
\includegraphics[width=\columnwidth]{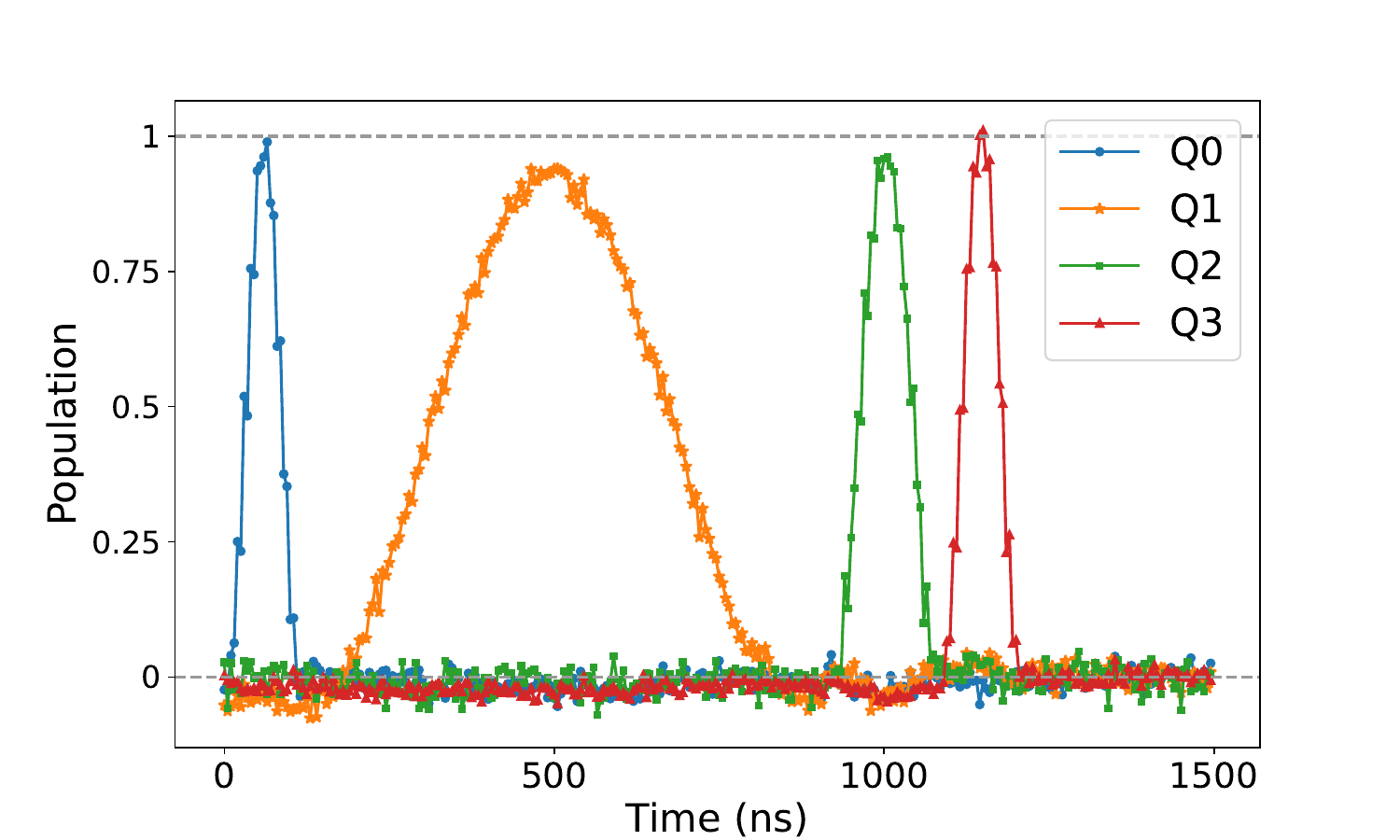}
	\end{center}
	\vspace{-0.6cm}
        \caption{\textbf{Qubit population measured with multiplexed readout.} An example of a pulse sequence where four transmon qubits are driven sequentially with 2 consecutive $\pi$-pulses. The qubit populations are sampled simultaneously at 5 ns intervals over the span of the gate sequence, with the readout handled by just 1 DAC and 1 ADC on the QICK board.} \label{fig:mux_readout_expt}
\end{figure}

The measured system consists of four capacitively coupled fixed-frequency transmon qubits. In this experiment, we needed simultaneous qubit readout for several purposes. First, it allowed us to track qubit populations over the time span of a sequence of gates that act on one or more qubits (e.g. Fig.~\ref{fig:mux_readout_expt}), which provides an easy-to-interpret visual representation of the success of a complex protocol. Second, simultaneous readout was necessary to perform quantum state tomography on sets of two or more qubits. Tomography requires simultaneous single-shot readout of all qubits involved in the tomography in order to categorize shots into the proper states.

In Fig.~\ref{fig:mux_readout_expt}, we show qubit measurements during a simple protocol in which we performed two consecutive $\pi$-pulses on each of the four qubits. At each point along the time axis, we measured the population of all qubits, where the y-axis is scaled by using the distance between the resonator I/Q peaks when its respective qubit is in $\ket{e}$ vs. $\ket{g}$.

\begin{figure}[!h]
	\begin{center}
\includegraphics[width=\columnwidth]{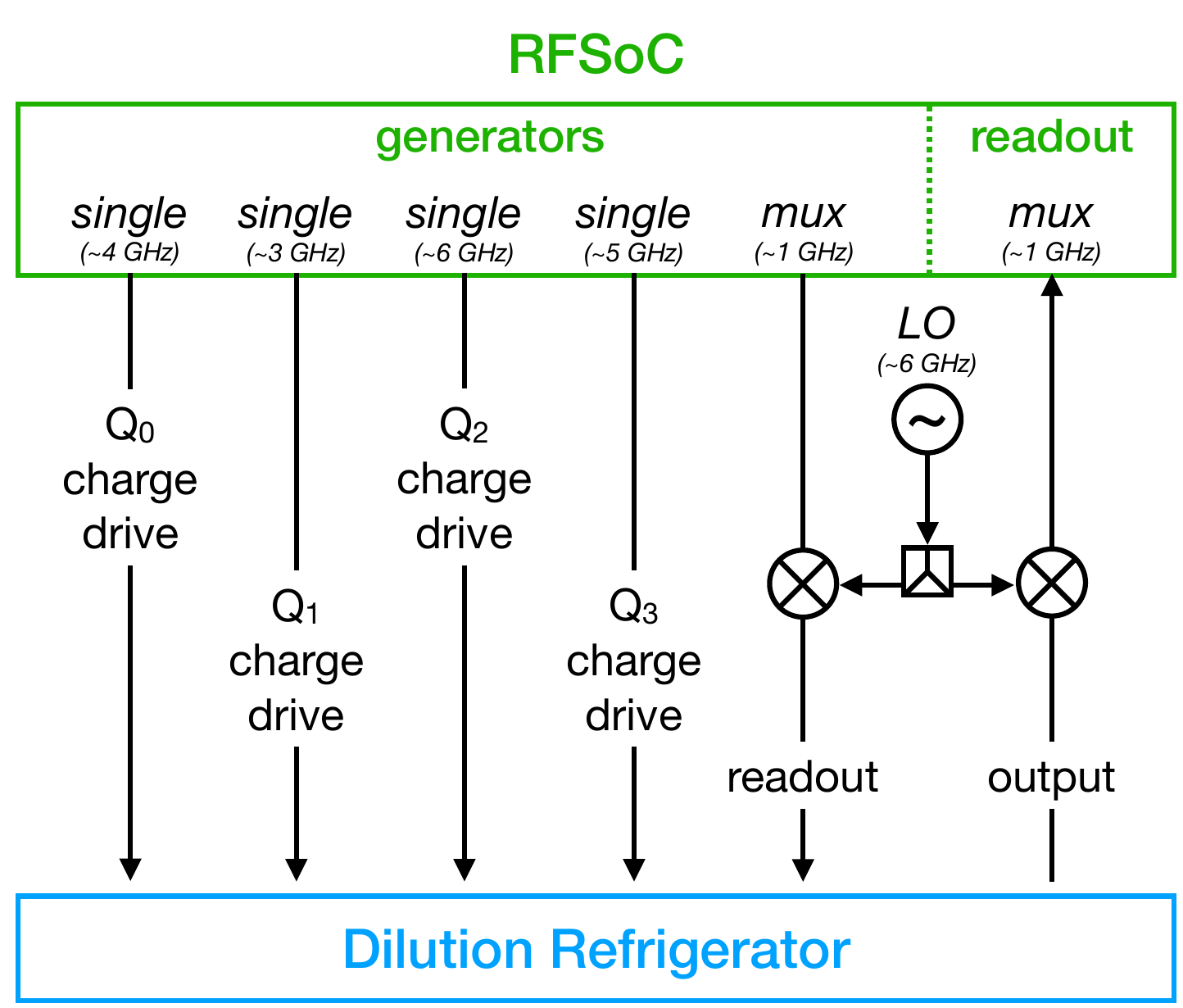}
	\end{center}
	\vspace{-0.6cm}
        \caption{\textbf{Wiring schematic for four-qubit simultaneous readout.} We use one full-speed generator per qubit to drive each qubit's charge line, plus one multiplexed generator and one multiplexed readout channel (mixed up/down with an external LO) to readout all four qubits simultaneously. We thus are able to perform simultaneous four qubit control and readout with just one RFSoC board. 
        } \label{fig:mux_readout_wiring}
\end{figure}

The room temperature measurement setup used to realize the previous measurement is shown in Fig.~\ref{fig:mux_readout_wiring}. Importantly, we were able to independently control each of the four qubits and perform simultaneous readout using just five DAC channels and one ADC, all on one QICK board, plus mixing with one external LO. The five DAC channels were respectively four standard generators used to drive each qubit and one mux generator used to measure all of them. In theory each of our four resonators could have been measured with just the RFSoC board alone. However, there were a few constraints that we needed to satisfy that led us to introduce an external mixer and LO: (1) The physical resonator frequencies were 6805, 5791, 7697, and 6966 MHz for qubits 0-3 respectively, which spanned a range of 1904 MHz. (2) The mux generator sampled at $f_{DAC}=6881.28$~MHz, so each of the four DDS's had a range of $f_{DAC}/4=1720$ MHz - less than the span of our resonator frequencies. The outputted frequencies were specified as $f_{mix}+[f_0, f_1, f_2, f_3]$, with each $f_{mix}+f_i$ limited by $f_{DAC}$, and each $f_i$ had to fall between $(-f_{DAC}/8, f_{DAC}/8)$. (3) The MUX readout (sampling at $f_{ADC}=2457.6$~MHz) required that each of the four frequencies fell in a different frequency bin of width $f_{ADC}/16=153.6$~MHz. To satisfy all of these constraints, we decided to set the RFSoC output at lower frequencies (950 + [-70, -816, 822, 91] MHz) and mixed it up with an LO (5925 MHz) from a Signalcore 5511A. This choice of mux and LO frequencies allowed us to span all of our readout frequencies by taking the positive sideband from the external mixer for qubits 0, 2, and 3 and the negative sideband for qubit 1. The ability to use the mux generator and readout thus significantly reduced the quantity of wiring and splitters needed, though it also reduced our flexibility in designing the filtering and amplification wiring for each readout frequency individually.

\section{Pre-distorted fast flux pulses}\label{sec:expt_pre-distorted}

In the Houck lab at Princeton, four frequency-tunable transmon qubits coupled in a ring configuration were prepared to explore the dynamics of particles with and without a synthetic magnetic field~\cite{Martinez2023}. Here, qubits play the roles of lattice sites and microwave excitations play the role of particles modeling the Bose-Hubbard Hamiltonian. The configuration of qubits is of a plaquette of a lattice whose energy bands contain one flat band. Under the addition of a synthetic magnetic field, all bands of the lattice become flat and all single-particle dynamics becomes localized. This is due to destructive interference arising from a combination of the Aharonov-Bohm effect and a particular lattice geometry ~\cite{VidalABCage}. One experimental challenge is that flat-band states are highly sensitive to disorder in the lattice site (qubit) frequencies. Therefore, any time-varying shifts in qubit frequency need to be minimized. 

The experimental sequence used to characterize dynamics consists of three steps: state initialization, time evolution, and readout. In each step, the qubits are biased to different frequencies by threading magnetic flux through the Josephson junction loop, modifying the qubit’s effective inductance. For state initialization, qubits are detuned from each other to be individually addressable. After initializing individual qubits to their excited states, the qubits are diabatically tuned onto resonance to start the time evolution. We require that this frequency ramp happens much faster than a characteristic tunneling time and that the qubit frequency is stable for the duration of the time evolution of a few microseconds. Finally, the qubits are detuned, freezing the dynamics, and the states of each qubit are measured using a dispersively coupled cavity.

\begin{figure}[ht!]
\begin{center}
\includegraphics[width= 1\columnwidth]{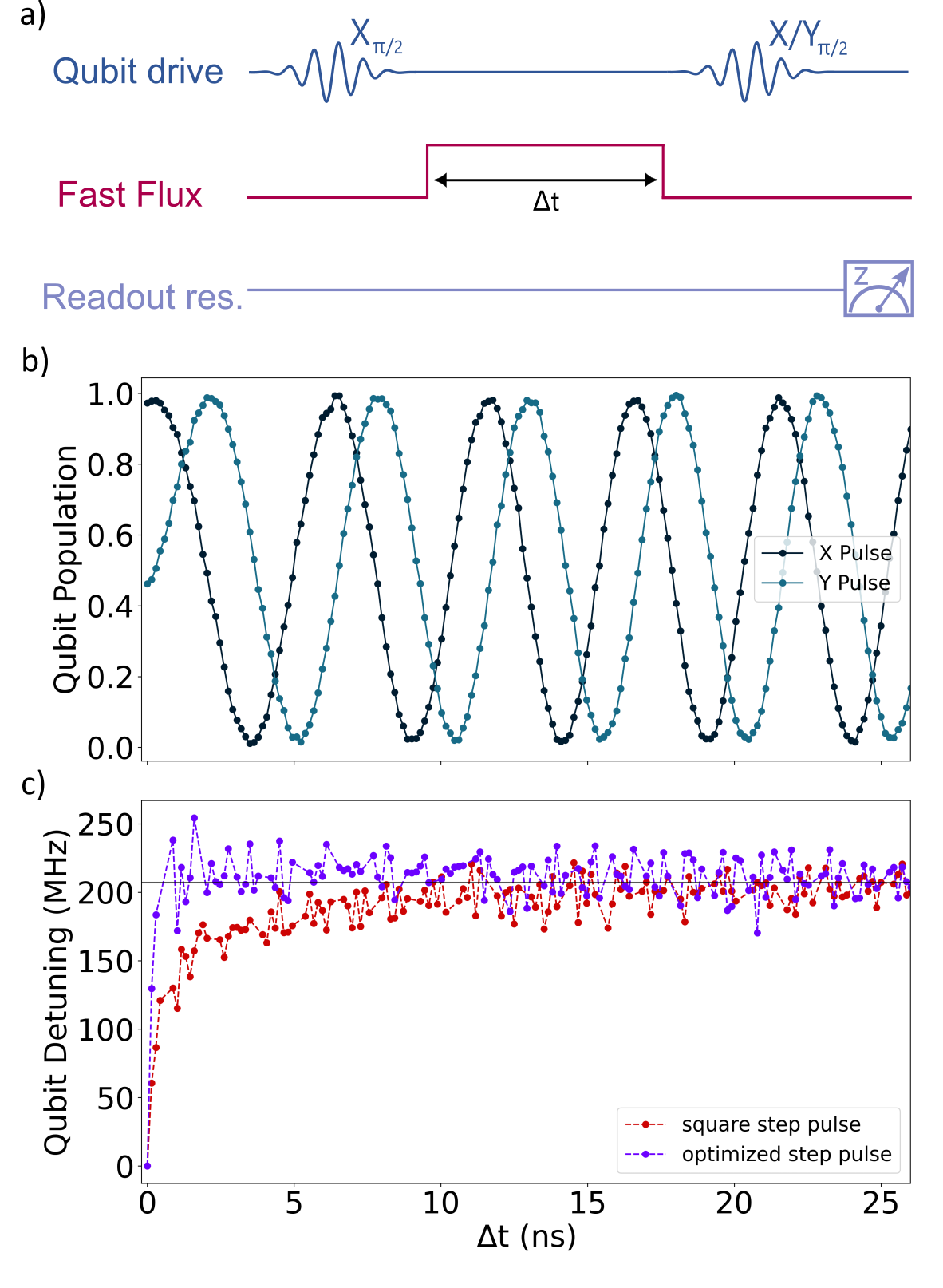}
        \caption{\textbf{Procedure for calibrating for short-time step-pulse distortions.} (a) Pulse sequence consisting of a step-pulse of variable length between two $\pi/2$ pulses. During the fast flux pulse, the qubit freqeuncy is detuned relative to the drive frequency and accumulates a relative phase. (b) Measured qubit population after applying the second $X_{\pi/2}$ pulse (black) or $Y_{\pi/2}$ pulse (blue). Rate of phase accumulation maps directly to the detuning between the qubit and drive. (c) Detuning between qubit and drive, inferred from (b), for a square step-pulse (red) and optimized step-pulse post-calibration (purple).}
        \label{fig:T2RamseyCalib}
\end{center}
\end{figure}

Ideally, the qubit frequency is changed instantaneously by sending a step-function pulse down the control lines in the dilution fridge. However, the shape of the fast-flux pulses generated at room temperature will be distorted by imperfections such as reflections and frequency-dependent attenuation as it travels down the fridge lines. On chip, the square step-function pulse arrives misshapen, greatly slowing down the frequency tuning of the qubit. These distortions can be corrected by ``pre-distorting'' the pulse shape on the control hardware such that the qubit receives a sharp and flat flux pulse. The ZCU216 board's RF-DAC outputs running at 6.88 GS/s were used to calibrate and precisely specify ``pre-distorted'' fast flux pulses with 145 ps resolution. 

The fast-flux pulses experience distortions on different time scales that are addressed individually. First, the qubit frequencies need to be shifted much faster than the characteristic tunneling time of qubits. For typical coupling rates of tens of MHz, qubits must be placed on resonance within a few nanoseconds. The second time scale requires the qubit frequency to remain constant during the evolution time ($\mu$s time scale) of the experiment. We compensate for both using different schemes.

For the short-time distortions, we use an experiment inspired by T$_2$ Ramsey to infer the frequency shift of the qubit during the fast flux pulse similarly to that done in Ref.~\cite{RamseyCalFF}. The pulse sequence is given in Fig.~\ref{fig:T2RamseyCalib}(a). We first initialize the qubit in the superposition state $\frac{1}{2}(\ket{0} + \ket{1})$ with a $X_{\pi/2}$ pulse. We then apply a fast flux pulse which detunes the qubit relative to the drive frequency. The qubit accumulates a phase, $\phi$, which we can measure by applying a second $X_{\pi/2}$ or $Y_{\pi/2}$ pulse before measuring the qubit population:
\begin{equation}
    e^{i\phi(t)} = \frac{X(t) + iY(t)}{\sqrt{X^2 + Y^2}}
    \label{Eq:phase_exponential},
\end{equation}
in which $\phi$ is directly proportional to the detuning, $\Delta$, between the qubit and drive:
\begin{equation}
    \phi(t) = \int_0^t \Delta(\tau)d\tau
\end{equation}

\begin{figure}[hb!]
\begin{center}
\includegraphics[width= 1\columnwidth]{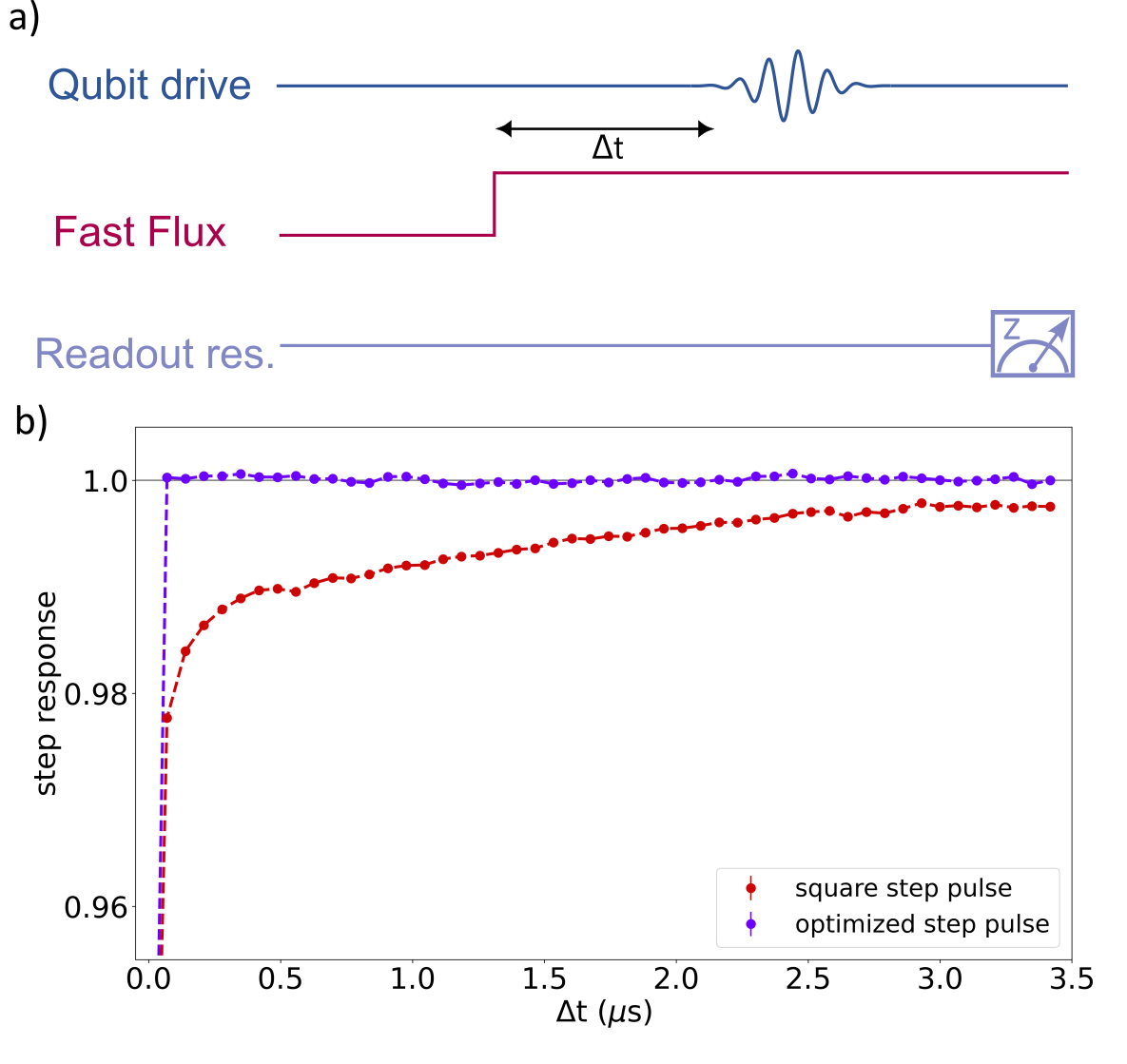}
        \caption{\textbf{Procedure for calibrating for long-time step-pulse distortions.} (a) Pulse sequence consisting of a fast-flux step-pulse, a qubit drive pulse with variable delay, and measurement of the qubit state via the coupled resonator. By sweeping the qubit drive frequency, the qubit frequency can be fitted for different delay times. (b) Qubit response to a square step-pulse (red) and optimized step-pulse (purple). Step response of 1 is normalized by dividing the fitted and target qubit frequencies.}
        \label{fig:SpecCalib}
\end{center}
\end{figure}

\begin{figure*}[t!]
\begin{center}
\includegraphics[width=0.99\textwidth]
{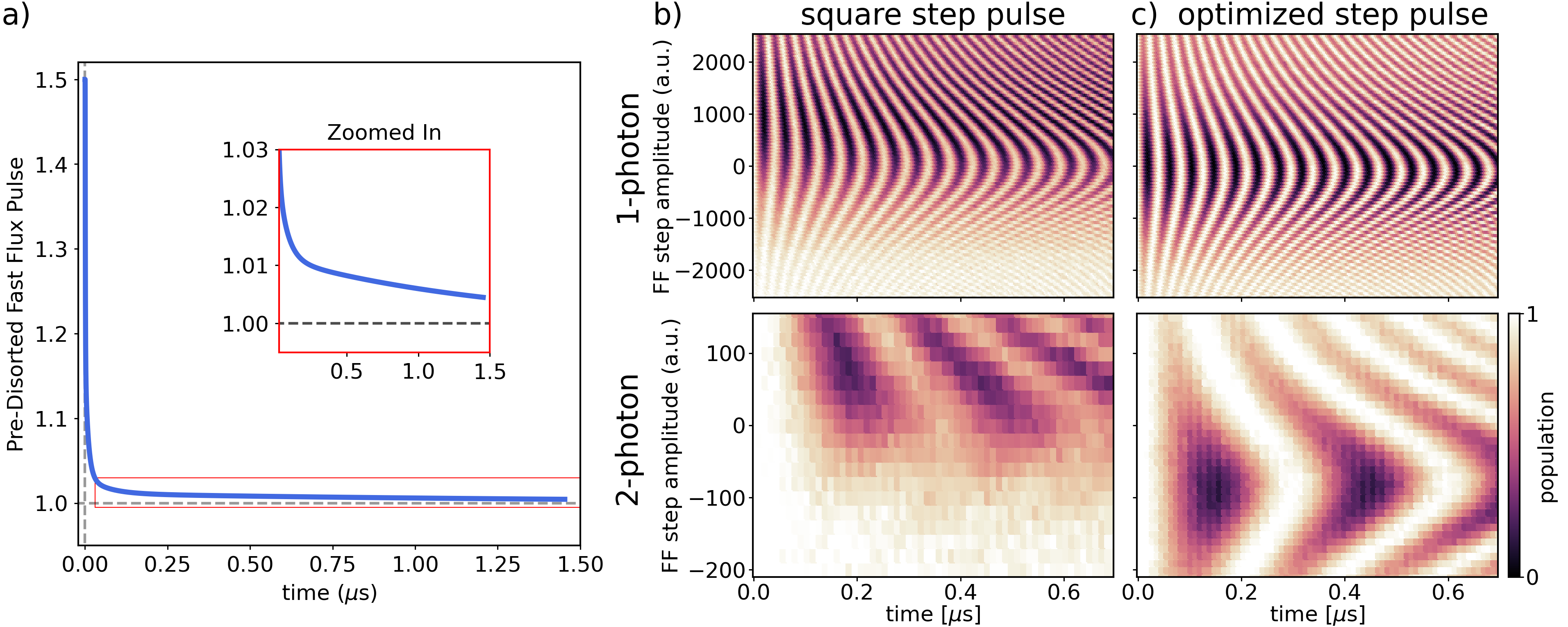}
        \caption{\textbf{Optimized step pulse and two-qubit swaps} (a) Amplitude of the pre-distorted step-pulse amplitude that results in a more optimized step response when reaching the qubit, normalized by a square step-pulse amplitude. Inset shows a zoom-in of the amplitude dominated by long-time distortions.  (b) (upper) Swapping of qubit excitation between two qubits, where the one qubit is initialized in the first excited state. After initialization, a step function is applied to the fast-flux line of that qubit with varying amplitude (“FF step amplitude”, y-axis) to bring it onto resonance with the second qubit that is biased there. Pre-calibration, the swaps are not complete and they are not symmetric with respect to the sign of the FF step amplitude. (lower) Same, but for a swap of the second excited state. (c) Same as (b), but with the calibrated fast-flux pulse. Fig.~modified from the supplement of Ref.~\cite{Martinez2023} with the permission of the authors.}
        \label{fig:extfig_ff}
\end{center}
\end{figure*}

By differentiating Eq.~\ref{Eq:phase_exponential}, we can measure $\Delta(t)$ and map out the qubit frequency response. In this scheme, we note that the initial qubit frequency is biased to a ``sweet spot” where the qubit frequency is first order insensitive to external flux. This is done to suppress any distortions when the qubit returns to the initial frequency that could alter the accumulated phase. The measured qubit population after the $X_{\pi/2}$ or $Y_{\pi/2}$ pulses is seen in Fig.~\ref{fig:T2RamseyCalib}(b). We convert the slope at each time step into the detuning between the qubit and drive frequency, plotted in red in Fig.~\ref{fig:T2RamseyCalib}(c). This data is then fit to a sum of exponentials to generate a transfer function, which we invert to define a pre-distorted pulse. The qubit response to the pre-distorted step pulse is shown in Fig.~\ref{fig:T2RamseyCalib}(c) where it approaches the target frequency within a nanosecond.

To calibrate for long-time distortions, we do a spectroscopy experiment similar to that in the supplement of Ref.~\cite{MaMott}. We send a step pulse on the flux line and then send a short Gaussian pulse with $\sigma$ = 7 ns to measure the qubit frequency for variable delay times. The amplitude is weaker than that needed to drive a $\pi$ pulse. In comparison to the previous scheme, this method measures the qubit frequency with higher precision but suffers from poor time resolution limited by the qubit frequency drifting while the spectroscopy drive pulse is applied. The pulse sequence is shown in Fig.~\ref{fig:SpecCalib}(a). We take spectroscopy data after different delay times and fit each time step to a Lorentzian function. The fitted qubit response to the square step pulse is seen in Fig.~\ref{fig:SpecCalib}(b). The data is similarly fit to a sum of exponentials and inverted to define the pre-distorted step-pulse. We find that the pre-distorted pulse allows for the qubit frequency to be stable within 0.1\% for the duration of the experiment.

Using the calibration from these two schemes, we define a pre-distorted pulse that is a sum of four exponentials, where the short time and long time calibrations each define two of these terms. We find that including additional terms did not noticeably improve the pulse performance. The amplitude of the pulse normalized to the amplitude of a square step pulse is shown in Fig.~\ref{fig:extfig_ff}(a). The pre-distorted pulse overshoots the final amplitude to rapidly bring the qubit onto resonance; this amplitude remains slightly elevated for long times which stabilizes the qubit frequency for the duration of the experiment. 

We evaluated the calibrated pulses by running a time evolution experiment between two qubits. Starting with the qubits detuned from each other, we initialized one qubit on either the first or second excited state. The qubits were then placed onto resonance using the fast flux pulses. Resonant qubit dynamics are characterized by full contrast oscillations:  the excitations ``swap'' between the qubits at a rate given by their coupling strength with full population transfer. When qubits are detuned from each other, the frequency of swaps increases but the excitations do not have complete state transfer. 


In Fig.~\ref{fig:extfig_ff}(b) and (c), we plot the population of the qubit with an initial excitation as a function of a fast-flux amplitude during the time evolution section of the experiment. We first detune one of the qubits by 200 MHz for state initialization. Using fast-flux pulses, we step the qubit to different fast-flux amplitudes shown in the y-axis for varying times. We note that we biased the qubits frequencies to be resonant when the fast-flux amplitude is close to zero. 
The upper panels show the swaps when a single excitation is initialized on one qubit, while the lower panels has two excitations initialized on one qubit. Two excitations swap via a second-order process yielding a significantly slower oscillation frequency.
The swaps using the square step pulse (Fig.~\ref{fig:extfig_ff}(b)) show that the slowest frequency oscillations--- when qubit frequencies are resonant--- vary as a function of time, indicating frequency drifts during the experiment. Additionally, the maximum population contrast does not overlap with the slowest frequency oscillations, indicating that the qubit does not reach the target frequency quickly enough. These discrepancies are significantly improved with the optimized step pulse (Fig.~\ref{fig:extfig_ff}(c)). 

The 145 ps time resolution of the QICK allowed us to characterize the time distortions in fast flux pulses necessary for state initialization as well as starting and freezing dynamics of multiple qubit systems. Using this analysis, we compensated for these distortions using arbitrary waveform generation enabled with the board. The high temporal resolution of QICK has the bandwidth to compensate for experiments with larger coupling rates and stricter requirements which will be crucial in our future experiments. Finally, we note that one section of the project required measuring correlations between multiple qubits. Doing so utilized the multiplex readout enabled by QICK (Sec~\ref{sec:multiplexed}) to measure the states of two qubits simultanously. 

\section{Phase-coherent parametric quantum operation}
\label{sec:expt_ParametricMeasurement}
\newcommand{\sg}{\text{\textit{i}SWAP}}

Parametrically induced interactions are widely used for constructing quantum gates in superconducting quantum systems. In these systems, a central non-linear coupler is typically coupled to one or more logical quantum modes such as qubits or cavities ~\cite{paraoanu2006microwave, reagor2018demonstration, Zhou2023, chapman2022high, goss2022high}. The desired gate dynamics between logical modes are activated by pumping the coupler mode at specific off-resonance frequencies that create the desired Hamiltonian from a higher-order (typically third- or fourth-order nonlinearity). Such parametric processes have the advantage that they can be easily applied among multiple fixed-frequency modes, with the frequency of the drive determining which parametric interaction is activated. However, during the parametric interaction, the parametric pumps also imprint extra  phases onto the quantum states,  similarly to the manner in which changing the phase of a resonant qubit drive can alter the angle over which the qubit is rotated. Consequently, most parametric gates possess extra phase elements tied to the phase of the parametric pump, which places strong phase coherence requirements on the control electronics that generate the pumps.

\begin{figure}[ht!]
\begin{center}
\includegraphics[width= 1\columnwidth]{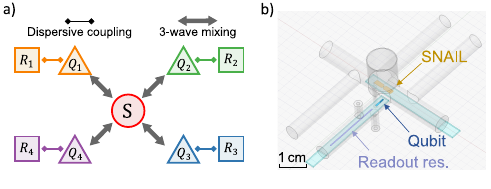}
\caption{\textbf{Coupling scheme and device schematic of a parametrically controlled 4-qubit module.} (a) Coupling scheme of the 4-qubit module. A central SNAIL mode ($S$) is coupled to four transmon qubit modes $Q_1-Q_4$, each qubit is also dispersively coupled to a readout resonator mode $R_1-R_4$. (b) Device schematic. A central aluminum tube for hosting a SNAIL chip is connected to four perpendicular channels, each is designed to host one qubit chip with strip line readout resonators. In the parametric measurement experiment described in the text, only one qubit chip and a SNAIL chip were used.}
\label{fig:hatlab_device}
\end{center}
\end{figure}

The Hatlab at the University of Pittsburgh has been conducting experiments on parametrically driven quantum state routers~\cite{Zhou2023} and quantum modules~\cite{mckinney2023co, mckinney2023parallel}.  Figure~\ref{fig:hatlab_device}(a) shows the coupling schematic of our 4-qubit quantum module. This device consists of four transmon qubits ($Q_1-Q_4$, with associated photon annihilation operators $q_{1-4}$) that are coupled to a central Superconducting Nonlinear Asymmetric Inductive eLement (SNAIL) coupler ($S$, with annihilation operator $s$)~\cite{frattini20173}. The SNAIL mode can be flux biased to a point with a third-order nonlinearity and negligible fourth-order nonlinearity, providing us with the initial non-linear term in the system Hamiltonian: $g_3(s+s^\dagger)^3$, where $g_3$ is the strength of the SNAIL's third order nonlinearity. As a result of the dispersive coupling between the SNAIL mode and each qubit mode, the third order non-linearity is shared among the qubits, leading to all possible third order terms of the three modes.  For the purposes of creating parametric exchange gates, we focus on the subset of terms which form the interaction Hamiltonian
\begin{equation}
    H_{\mathrm{int}} / \hslash = \sum_{i\neq j}  {g_{ijs} (q_i^\dagger q_j s^\dagger + q_i q_j^\dagger s)},
\end{equation}
where $g_{ijs}$ is the 3-wave-mixing coefficient among qubit modes $i, j$ and the SNAIL mode. Based on these terms, we can selectively activate the desired exchange interaction between two qubit modes, say $Q_1$ and $Q_2$, by pumping the SNAIL mode at the frequency $f_p = f_2-f_1 +\delta$, where $f_i$ is the mode frequency of $Q_i$, $\delta$ is the pump detuning added due to the relative AC Stark shift between qubits 1 and 2, and we assume $f_2>f_1$. For a given pump phase $\phi_p$, the effective Hamiltonian can be written as:
\begin{equation}
    H_{\mathrm{eff}}/\hslash = g_2^{\mathrm{eff}} (q_i^\dagger q_j e^{i\phi_p} + q_i q_j^\dagger e^{-i\phi_p}),
\label{eq:exchange_interaction}
\end{equation}
where $g_2^{\mathrm{eff}}=g_{ijs} \sqrt{n_p}$ is the effective exchange rate and $n_p$ is the pump strength expressed in units of photons.  Such exchange interaction between two qubits creates a continuous family of gates, described $\sg^\alpha$, which can be used to implement a variety of universal two-qubit gates in quantum circuits. 

In the above gate operation, the impact of the pump phase can be understood by considering the following scenario: $Q_1$ is initially prepared to a superposition state with initial phase $\phi_1$, while $Q_2$ is in the ground state, i.e. $\ket{\psi_0} =  1/\sqrt{2} (\ket{g} + e^{i\phi_1} \ket{e}) \bigotimes \ket{g}$. After a complete  \sg ~gate with pump phase $\phi_p$, the final state of the qubits will become:  $\ket{\psi} = \ket{g} \bigotimes 1/\sqrt{2} (\ket{g} + e^{i(\phi_1+\phi_p)} \ket{e})$. To check the phase of the resulting state, we apply a $\pi/2$-pulse with phase $\phi_2$ on $Q_2$ and measure its $\sigma_z$ expectation value. The measurement result will then be $\expval{\sigma_{z,2}} = -\cos{(\phi_1 + \phi_{p}-\phi_2)}$. Note that each of the three phases involved here are originally carried by the three microwave drive channels at frequencies $f_1$, $f_p$, and $f_2$, with the qubit drive frequencies  $f_1, f_2 \approx 3-6 \text{GHz}$, and the pump frequency $f_p \approx 0.5-2 \text{GHz}$. Thus,  to ensure that the measurement results stay the same between repetitions; the three drive channels must stay phase coherent with each other under the constraint 
\begin{equation}
    [\phi_1 (t) + \phi_p (t) - \phi_2 (t)]|_{t=t_N}  = constant,
\label{eq:phase_coherence}
\end{equation}
where $t_N$ is the start time of experiment repetition $N$. 

\begin{figure}[ht!]
\begin{center}
\includegraphics[width= 1\columnwidth]{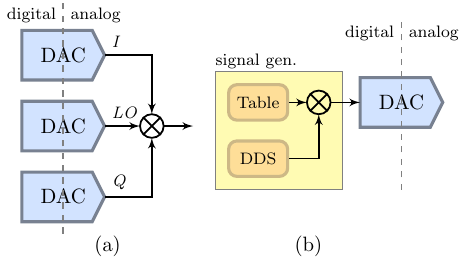}
\caption{\textbf{Typical circuits for synthesising a microwave drive for superconducting qubit control.} The possible common subsequent components (e.g. amplifiers, dividers) are omitted. (a) Analog mixer based up-conversion circuit. Two DAC channels are connected to the I and Q ports of an analog mixer, whose local oscillator is provided using a continuous signal generator with a third DAC output. (b) Direct pulse generation with DDS. Complex IQ mixing is performed in the digital domain, and one DAC output provides the required pulse. Here, only an extra low- or band- pass filter is needed after the DDS DAC to filter out the image tones in other Nyquist zones.}
\label{fig:hatlab_drive_circuit}
\end{center}
\end{figure}

This constraint can be satisfied with different methods, depending on the microwave control circuit setup being used. Fig.~\ref{fig:hatlab_drive_circuit}(a) shows an analog up-conversion circuit that is typically used to generate control pulses for superconducting qubits. In this setup, two output channels of a relatively low frequency AWG (typically with bandwidth of $\sim 500 \text{MHz}$) are connected to the I and Q ports of an analog IQ mixer, whose local oscillator (LO) is provided by a high frequency signal generator. In this setup, the phase of each AWG pulse can be set to arbitrary value $\phi^I$, while the generator phase is continuously evolving and cannot be reset in real time. The total phase of each drive channel now comes from two parts, i.e.
\begin{equation}
    \phi(t) = 2\pi * \int_{-\infty}^t f^L(t) \,dt\ + \phi^I,
\end{equation}
where $f^{L(t)}$ is the frequency of the signal generator. To satisfy the phase coherence constraint in Eq.~\ref{eq:phase_coherence} for ideal, stable generators, we can choose the generator frequencies ($f_i^L$) such that $f_1^L+f_p^L-f_2^L = 0$ to remove the absolute time dependent part of the total phase, and then choose the AWG pulse frequencies ($f_i^I$) accordingly, such that $f_i^I = f_i - f_i^L$. However, for experiments that requires averaging over long time periods, this constraint places a strong requirement on the relative stability between all the signal generators and the AWG channels, which may be challenging for conventional microwave electronics. 

Alternately, QICK offers a much simplified way to generate the control pulses. As in Fig.\ref{fig:hatlab_drive_circuit}(b), the pulse for each control channel can be generated using DDS, eliminating the need for additional analog mixer and generator hardware. Crucially, the QICK firmware offers the functionality of resetting the phase of all the DDS channels simultaneously at well-defined times. Thus, the absolute time dependent parts of the drive phases in Eq.~\ref{eq:phase_coherence} are completely removed, which makes the phase coherence condition automatically satisfied. Since all the drive phases and frequencies are defined digitally on a single FPGA chip, even if the absolute frequency of each channel drifts over time, their relative phases will always stay locked, which also makes long-term relative phase stability easily realizable. Furthermore, as needed, one can directly source a LO channel using the ZCU216 board, enabling coherent reset. 

To have full control over the device described in Fig.~\ref{fig:hatlab_device}(a), we will need one qubit drive channel plus one readout resonator drive channel for each qubit, and there could be 6 possible SNAIL drive frequencies for the parametric two-qubit gates. With the limited instantaneous bandwidth available in low frequency AWGs (as in Fig.~\ref{fig:hatlab_drive_circuit}(a)), all these drive channels will need to be synthesised individually, which requires a total of 14 signal generators plus IQ mixers, and 28 AWG channels. With a high-frequency DDS, the whole device can be controlled with only 14 DAC channels on a single ZCU216 board. This can be even further simplified with frequency-multiplexed digital generators. Therefore, the DDS pulse generation provided by the QICK-controlled RFSoC boards not only provides stable phase coherent control, but also significantly reduces the complexity of microwave hardware setup.

As an example demonstration of the phase stability of QICK-controlled parametric quantum operation, we present the results from a parametric readout experiment conducted in a subset of our 4 qubit module. For this particular experiment, only one qubit and a SNAIL are used, as depicted in Fig.~\ref{fig:hatlab_device}(b).

The parametric measurement process for this experiment is activated by applying two microwave pumps on the SNAIL mode at the difference and sum frequencies between the SNAIL and qubit mode. The difference frequency pump will activate the photon exchange (conversion) interactions between the qubit and SNAIL mode, similar to the \sg~ interaction described earlier with Eq.~\ref{eq:exchange_interaction}. The sum frequency pump will turn on the two-photon transition (gain) process~\cite{poletto2012entanglement} that, instead of exchanging photons, will jointly excite or destroy photons in the two modes. By turning on both pumps simultaneously and tuning their relative amplitudes to have matched interaction strength, we can activate the effective Hamiltonian:

\begin{align}
\begin{split}
    H_{\mathrm{eff}}/\hslash &= g(q^\dagger s e^{i\phi_c} + q s^\dagger e^{-i\phi_c} + q^\dagger s^\dagger e^{-i\phi_g} + q s e^{i\phi_g})\\ 
    &= g(\cos{(\phi_m)} \sigma_x - \sin{(\phi_m)} \sigma_y)(s^\dagger e^{-i\phi_s} + s e^{i\phi_s})
\end{split}    
\end{align}
where $\phi_g$ and $\phi_c$ are the phases of the gain and conversion pumps respectively, and $\phi_m = (\phi_g-\phi_c)/2, \phi_s = (\phi_g+\phi_c)/2$. By controlling the two pump phases $\phi_g$ and $\phi_c$, the qubit can be projected along an arbitrary direction on the XY plane of the Bloch sphere, and creates a qubit state dependent coherent state in the SNAIL mode. This process effectively performs a measurement on the qubit along the projection axis, and the measured result can be collected by demodulating the signal coming out from the SNAIL port.

\begin{figure}[ht!]
\begin{center}
\includegraphics[width= 1\columnwidth]{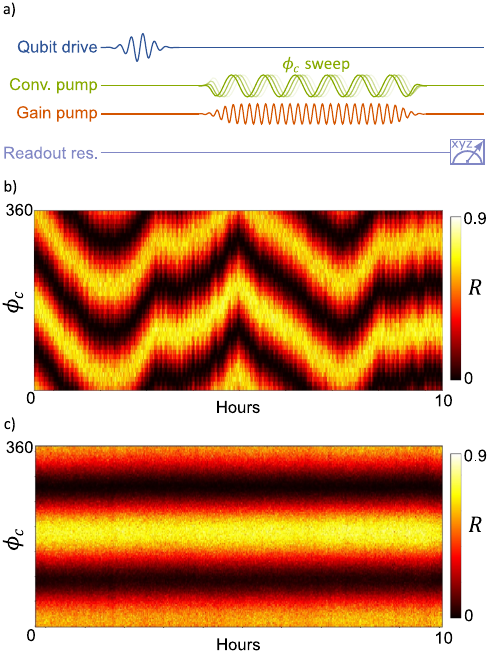}
\caption{\textbf{Pulse sequence and longtime phase stability data for calibrating the measurement direction in the parametric measurement experiment.} (a) Pulse sequence for parametric measurement direction calibration. The qubit is first prepared in a $\ket{X_+}$ state with a $\pi/2$-pulse, then the gain and conversion pumps are applied simultaneously on the SNAIL mode to perform the parametric measurement. Finally, conventional state tomography is performed on the qubit using the readout resonator. The phase of the conversion pump ($\phi_c$) is swept to align the parametric measurement direction with the $\ket{X_+}$ direction. (b) Phase stability data obtained from running the $\phi_c$ sweep experiment using a mixer-based control circuit for a duration of 10 hours. The phase exhibits a 6 minute period drift caused by imperfections in the phase locked loop of the signal generators, and also shows a larger overall phase drift over several hours due to temperature fluctuations. In contrast, (c) show the same phase stability test performed using a QICK-controlled RFSoC board in which the phase remained stable throughout the entire 10-hour experiment.}
\label{fig:hatlab_x_msmt}
\end{center}
\end{figure}

Fig.~\ref{fig:hatlab_x_msmt}(a) shows the pulse sequence used to calibrate the direction of the parametric measurement. The qubit is first prepared to the $\ket{X_+}=(\ket{g}+\ket{e})/\sqrt{2}$, then the gain and conversion pumps are applied simultaneously. After the pumps are off, a full qubit state tomography is performed using the readout resonator. When the parametric measurement direction is aligned with the $\ket{X_+}$ or $\ket{X_-}=(\ket{g}-\ket{e})/\sqrt{2}$ direction, the length of the qubit Bloch vector $R=\sqrt{\expval{\sigma_x}^2+\expval{\sigma_y}^2+\expval{\sigma_z}^2}$ will be preserved during the parametric measurement. On the other hand, when the qubit is prepared in the perpendicular states $\ket{Y_\pm}=(\ket{g}\pm i \ket{e})/\sqrt{2}$, the qubit will collapse to the mixed state $1/2\ket{X_+}\bra{X_+}+1/2\ket{X_-}\bra{X_-}$ with $R=0$ after the measurement. The measurement direction is simply calibrated by sweeping the phase of one of the pumps, say $\phi_c$. Similarly to the \sg~ case, here the qubit drive and both pump channels must stay phase coherent under a slightly different constraint:
\begin{equation}
        [\phi_q (t) - (\phi_g (t) - \phi_c (t))/2]|_{t=t_N}  = constant,
\label{eq:phase_coherence2}
\end{equation}
where $\phi_q$ is the phase of the qubit drive pulse. Moreover, the relative phase must stay stable over a long time for experiments that requires a large number of average repetitions, e.g. measurement efficiency characterization~\cite{hatridge2013quantum}. Fig.~\ref{fig:hatlab_x_msmt} (b) shows the result of running the pump phase calibration experiment for 10 hours using an analog mixer-based setup. Although all the signal generators and AWG channels used were locked to an external $10~\text{MHz}$ Rubidium clock, we still observed  a $\sim 6$ minute period  phase drift due to the imperfections of the phase-lock-loop in the signal generators,  as well as an overall slow drift we attribute to changes in room temperatures over time~\cite{ball2016role, Xia2023SubHarmonic}. Such drift makes all experimental studies extremely tedious, as re-calibration is required every few minutes. More importantly,  as the advancements of longer coherence qubits~\cite{Place2021,ding2023high, Zhang2023, ganjam2023surpassing} and the emergence of novel quantum error correction schemes~\cite{sivak2023real, Krinner2022, chou2023demonstrating} gradually allowing the execution of longer quantum circuits, it is becoming more crucial that the phase drifts in classical control electronics do not set a limit on the duration of quantum circuits we aim to execute. In contrast, Fig.~\ref{fig:hatlab_x_msmt} (c) shows the same experiment performed  using a QICK-controlled RFSoC board, where all the drives were generated using direct digital synthesis with phase reset on each repetition. The measurement phase remained stable over the entire 10-hour experiment, providing clear evidence of the relative phase stability of the RFSoC and its excellent suitability for parametric quantum operations.

In general, the relative phase coherence between different drive channels is essential for performing parametric quantum operations. The requirements for phase coherence can vary depending on the specific parametric process being implemented, and can be expressed as specific constraints on the relative phases between the drive channels (e.g., Eq.~\ref{eq:phase_coherence} and \ref{eq:phase_coherence2}). While it is possible to construct custom analog interferometer circuits with high-stability microwave generators to meet these constraints, using DDS-based pulse generation with a well-defined phase reset function is a more efficient and convenient solution. Because it offers precise control over the frequencies and phases of the generated signals, and also greatly reduces hardware requirements as well as the complexity of experimental setup. Note that the phase reset functionality available in the QICK firmware is vital here as it erases the historically accumulated phases, thereby eliminating the need for long-term high-precision phase stability in control electronics, which could be hard to realize.

\newcommand{\siswap}{\sqrt{i\mathrm{SWAP}}}
\newcommand{\sbswap}{\sqrt{b\mathrm{SWAP}}}

\section{Phase-sensitive parametric entangling gates}\label{sec:expt_ParametricEntangler}


\begin{figure}[ht!]
\begin{center}
\includegraphics[width= 1\columnwidth]{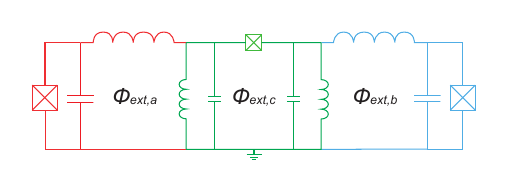}
\caption{\textbf{Circuit diagram of the inductively-coupled fluxonium qubits experiment}. Qubit A is in red, qubit B is in blue, and the tunable coupler is in green. We use long $(>200)$ chains of Josephson Junctions to realize the kinetic inductance in our qubit, and have shorter, shared chains to provide galvanic coupling between each individual fluxonium with the coupler. Each loop (for the qubits as well as for the coupler) has a dedicated flux bias line, for both DC biasing and low-frequency RF drives.
}
\label{fig:fluxonium_circuit}
\end{center}
\end{figure}

In a Schuster lab experiment conducted at the U. of Chicago, we used a tunable, fluxonium-like coupler~\cite{weiss2022fast} to galvanically couple two low-frequency, heavy-fluxonium qubits~\cite{Zhang2021, Zhang2023}. A key feature of our design is that for a particular DC bias, the tunable coupler has a zero-coupling ``off-position'', where the $\textit{XX}, \textit{ZZ}$ qubit-qubit coupling terms are nearly nulled. A circuit diagram is shown in Fig~\ref{fig:fluxonium_circuit}, and the effective Hamiltonian can be written as 

\begin{align}
\label{eq:effHam}
H_{\text{eff}} &= -\sum_{\mu=a,b}\frac{\omega_{\mu}}{2}\sigma_{z}^{\mu}-\Omega_{\mu}\sigma_{x}^{\mu} + J\sigma_{x}^{a}\sigma_{x}^{b}
+\zeta\sigma_{z}^{a}\sigma_{z}^{b},
\end{align}
where $\sigma_{x}^{\mu}, \sigma_{z}^{\mu}$
are the qubit Pauli operators in the basis of symmetric and anti-symmetric wavefunctions. When biased at the coupler ``off-position'', experimental data confirms that the residual $\textit{ZZ}$ couplings is less than $100$ Hz. At that DC bias, we dynamically activate the $\textit{XX}$ coupling by rf driving at either the sum or difference of the qubit frequencies. 


\begin{figure}[ht!]
\begin{center}
\includegraphics[width= 1\columnwidth]{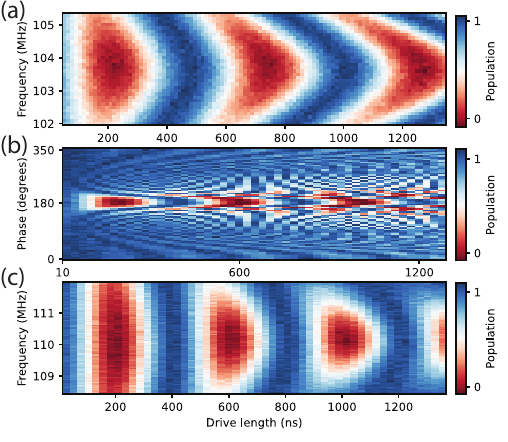}
\caption{\textbf{Compensation pulses showing a rf crosstalk calibration and reduction} (a) Rabi measurement with variable length pulses of the coupler drive, in the neighborhood of the sum frequency $\omega_a + \omega_b$ of the qubits. The colorscale here indicates the population present in qubit A. In the absence of crosstalk cancellation pulses, the maximum Rabi amplitude oscillation is $7$MHz detuned from the bare qubit frequencies due to a large AC-Stark shift. (b) The calibration sweep for one cancellation pulse, where the relative phase between the qubit drive and the coupler drive is adjusted. One can see the maximum contrast when the compensation pulse has roughly the opposite phase of the coupler drive. (c) Same Rabi measurement as (a), but with applied compensation pulses on both $\Phi_{\mathrm{ext}, a}, \Phi_{\mathrm{ext}, b}$. Note that the maximum Rabi amplitude oscillation is now centered at exactly $110.2$MHz, which is what we expect given our single qubit measurements. Data for fig.~modified from the supplement of Ref.~\cite{Zhang2023} with the permission of the authors.}
\label{fig:crosstalk_cancellation}
\end{center}
\end{figure}

When galvanically coupling our fluxonium qubits, we have both intrinsic crosstalk due to the construction of our circuit, as well as a significant geometric crosstalk as the qubit loops are adjacent to one another. We measure this DC crosstalk matrix to have up to 25$\%$ crosstalk between flux bias lines. To realize a parametric interaction between our qubits using the coupler, we perform an rf drive at the sum and difference of the qubit frequencies. However, this effectively induces a detuned AC-Stark shift on each individual qubit. Because the optimal drive amplitude of the parametric oscillation is dependent on the effective qubit frequencies, such AC-Stark shifts change the frequency of the coupler drive at which there is maximum Rabi amplitude contrast. From other single-qubit measurements, we determine that the bare frequencies of our qubits are $\omega_a = 48.4$ and $\omega_b = 61.8$ MHz, meaning that the optimal coupler drive for the bSWAP interaction~\cite{weiss2022fast} should be at $\omega_a + \omega_b = 110.2$ MHz. However, when performing a frequency-length Rabi experiment using the coupler drive, we see that the center of this Rabi chevron has been shifted down by $\approx 7$ MHz, see Fig~\ref{fig:crosstalk_cancellation}(a).

In general, such crosstalk can be cancelled using the additional degrees of freedom afforded by having individual flux lines for each qubit loop. We implement this correction with simultaneous flux crosstalk cancellation pulses, played at the same time as driving our coupler~\cite{Sheldon2016Procedure}. Along the $\Phi_{\mathrm{ext}, a}, \Phi_{\mathrm{ext}, b}$ bias lines, we use pulses with the same shape, frequency, and length as our two-qubit gate drive, but with calibrated amplitudes and relative phase offsets. Because these compensation parameters are independent to the coupler drive, we can measure the Rabi oscillation of the coupler while sweeping the phase and amplitude of our compensation pulses for each parameter of both qubit A and B's flux bias lines. We show the phase sweep of a cancellation pulse in Fig~\ref{fig:crosstalk_cancellation}(b), revealing that the optimal relative phase between the cancellation channels and the coupler channel is $\approx 180 \deg$. 

By applying the cancellation pulses and performing the same Rabi experiment as before, we see oscillations between $\ket{00}$ and $\ket{11}$ with maximum Rabi contrast at exactly the sum of qubit frequencies $\omega_a+\omega_b$, see Fig~\ref{fig:crosstalk_cancellation}(c). This indicates that there is no more off-resonant drive on the qubits, and that our compensation pulses effectively eliminated flux crosstalk from the coupler flux drive.


With our calibrated crosstalk cancellation pulses and our ability to realize generic $\textit{XX}$ interactions using the RF drive on our coupler, we can realize a $\ket{gg}\leftrightarrow\ket{ee}$ oscillation. The Hermitian matrix for such a gate can be written as

\begin{align}
\sqrt{\phi_b\mathrm{SWAP}}=
\begin{pmatrix}
\cos{\theta} & 0 & 0 & i e^{i\phi_D}\sin{\theta} \\
0 & e^{i\phi_{01}} & 0 & 0 \\
0 & 0 & e^{i\phi_{10}} & 0 \\
i e^{i(\phi_{11}-\phi_D)}\sin{\theta} & 0 & 0 & e^{i\phi_{11}}\cos{\theta}
\end{pmatrix}
\end{align}

where $\phi_D$ is the phase of the coupler drive, and $\phi_{01}$, $\phi_{10}$, $\phi_{11}$ are phases due to the frequency shift of levels while the drive is on, which have relationship $\phi_{11}=\phi_{01}+\phi_{10}+\phi_{zz}$. From this matrix, we can see that control of the coupler drive phase, along with the application of single qubit Z gates, allows us to exactly realize a $\sbswap$ gate. (A similar matrix can be written for the $\siswap$ oscillation). Furthermore, we recognize that both the phase of individual qubits and the phase of the coupler are set by the individual AWG channels. Thus, the relative phase between our channels directly leads to an effective rotation angle within the \textit{fSim} family of gates. Conversely, if there was noise in either the absolute or relative phases of our channels, we would not be able to realize high fidelity parametric gates.

\begin{figure}[t!]
\begin{center}
\includegraphics[width= 1\columnwidth]{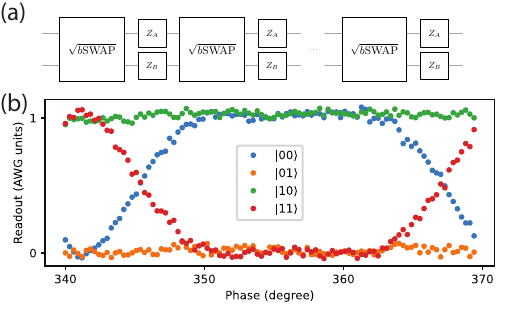}
\caption{\textbf{Calibration of coupler and single qubit phase in preparation of a $\sbswap$ gate} (a) The pulse sequence used for calibration. This sequence is constructed to measure the drive induced phase $\phi_{11}$, and each block is repeated 402 times in order to amplify small errors. (b) The measured phase relationship while sweeping $\phi_{A}$ phase gates ($Z_A$). Notice that this sequence amplifiers errors, such that a change of $<1\deg$ of $\phi_{A}$ results in a $10\%$ change in the state population at the maximally sensitive locations. We find that the optimal phase angle is at $357\deg$, corresponding to $\phi_{01} + \phi_{10} = 3 \deg$. Fig.~modified from the supplement of Ref.~\cite{Zhang2023} with the permission of the authors.}
\label{fig:phase_calib}
\end{center}
\end{figure}

In order to execute pure $\siswap$ and $\sbswap$ gates, we need consistent, individual control of the relative phases of the various AWG channels, which the QICK enables To determine the appropriate phases with very high accuracy, we amplify the error and make measurements. The sequence that is used to do this is shown in Fig.~\ref{fig:phase_calib}(a), inspired by similar protocols in Ref ~\cite{Ganzhorn2020Benchmarking}. We play a series of $\sbswap$ gates with single-qubit phase ($Z$) gates interleaved between each pair of them, and sweep the rotation angle of all the phase gates ($Z_A$) simultaneously. When we play $4n+2$ $\sbswap$ gates, we can only get a full $\pi$ rotation when $\phi_A+\phi_B=-\phi_{11}$. Thus, we measure the drive induced phase $\phi_{11}$ with a $\phi_A$ sweep, as shown in Fig.~\ref{fig:phase_calib}(b). Using similar error amplification sequences, we can also measure the drive induced phase $\phi_01$ and $\phi_10$. We compensate these phases by setting qubit A $Z$ gate phase $\phi_A=-\phi_{10}$, qubit B $Z$ gate phase $\phi_B=-\phi_{01}$, and obtain a $\sbswap$ gate with correct phases.

\begin{figure}[ht!]
\begin{center}
\includegraphics[width= 1\columnwidth]{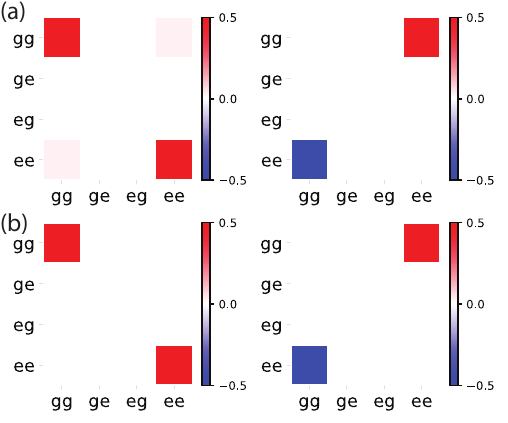}
\caption{\textbf{The density matrix of a Bell state prepared with our parametric gate, found with quantum state tomography (QST)} (a) We initialize our qubit to the $gg$ state and perform a single application of the $\sbswap$ gate. This prepares the Bell state, $(\ket{gg} + i \ket{ee})/\sqrt{2}$. Since performing QST requires averaging over thousands of measurements in different basis, a pure two-qubit density matrix implies a stable phase relation between QICK AWG channels. We measure a purity and state fidelity that are limited primarily by state preparation and measurement errors to be 95\%. (b) The density matrix for the ideal Bell state. Fig.~modified from the supplement of Ref.~\cite{Zhang2023} with the permission of the authors.}
\label{fig:state_tomo}
\end{center}
\end{figure}

After calibrating our $\sbswap$ gate, we use it to prepare a Bell state $(\ket{gg} + i \ket{ee})/\sqrt{2}$ (Fig.~\ref{fig:state_tomo}). To determine the state preparation fidelity, we choose to perform quantum state tomography (QST). This involves the measurement of our state in an overcomplete set of basis states, and subsequently performing maximal likelihood estimation to find the most likely corresponding density matrix (subject to physical constraints). In this experiment, we average 50000 times for measurement along each of the 9 basis vectors. If there were unstable relative phase between the QICK AWG channels, this would be revealed as a decrease in the purity of density matrix. In the worst-case scenario, where a completely random relative phase exists between control channels for each shot, this would result in a completely mixed density matrix. However, by using the \textit{phase-reset} firmware to synchronize the phases between the QICK AWG channels, we find a state purity of 95\%, indicating consistent phase preparation.


\begin{figure}[ht!]
\begin{center}
\includegraphics[width= 1\columnwidth]{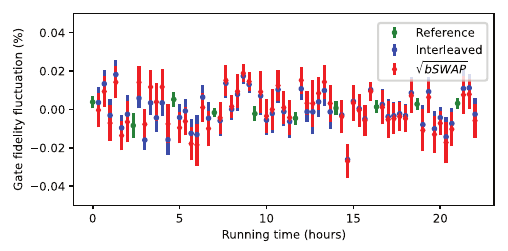}
\caption{\textbf{Variance in cross entropy benchmarking over 20 consecutive hours.} To benchmark our gate, we execute continuous cross entropy benchmarking sequences on our system. Each individual sequence is up to 600 layers of randomly-selected two-qubit Clifford gates, interleaved with the reference $\sbswap$ gate. For each cross-entropy benchmark sequence, we find an average fluctuation in gate fidelity to be $3 \sigma \approx 0.02\%$. Throughout the entire sequence, the average fidelity measured did not drift by more than $\pm 0.03 \%$. Data used for fig.~modified from Ref.~\cite{Zhang2023} with the permission of the authors.}
\label{fig:fluxonium_benchmarking}
\end{center}
\end{figure}

Using these compensation pulses, we can perform full benchmarking of the $\sbswap$ gate, with a continuous cross-entropy benchmarking result over the course of 25 hours. Here, we show the variance in gate fidelity over this period in Fig~\ref{fig:fluxonium_benchmarking}. During this period, no adjustments were made to any of the pulse parameters, nor to the crosstalk cancellation parameters. We know that our gate is sensitive to phase (see Fig.~\ref{fig:phase_calib}(b) and frequency (see Fig.~\ref{fig:crosstalk_cancellation}(c)), while amplitude errors would correspond to under- or over-rotation of the gate. Therefore, the consistency of the cross-entropy benchmarking results can be used as a metric to quantify the stability of these three parameters between our QICK AWG channels.

We find that not only does the cross-entropy benchmarking show a very high two-qubit gate fidelity, where the fidelity of the $\sbswap$ gate is greater than $99.9\%$, we find that there are very few fluctuations in the gate fidelity over the span of 25 hours. As shown in Fig.~\ref{fig:fluxonium_benchmarking}, each individual cross-entropy benchmarking sequence has a statistical fit infidelity on the order of $0.02\%$. Furthermore, there is no significant drift in the average gate fidelity, with a maximum fluctuation of no more than $0.03\%$. In our error analysis, we attribute the majority of the gate infidelity to qubit relaxation channels, not through coherent errors that can be attributed to the QICK AWG channels.

\section{Conclusion}\label{Conclusion}

The experimental use cases presented here show how the QICK can be applied to control tasks relevant to superconducting qubit systems. Multiplexed readout is used in virtually all scaled-up planar superconducting qubit systems, including those from academic labs~\cite{Krinner2022,Saxberg2022}. Mixer-free readout simplifies the control loop and can extend the control benefits gained from direct digital synthesis~\cite{Kalfus2021}. Pre-distorted fast flux pulses are useful for any control scheme that includes flux-tunable qubits or couplers~\cite{Zhang2021,Stehlik2021}. Parametric quantum operations are generally advantageous in qubit architectures as they simultaneously allow large coupling strengths and large qubit-drive detunings~\cite{Jin2023, Xia2023SubHarmonic}. 

Ongoing work on the QICK is branching into several directions. The QICK team has developed new lab tools running on the RFSoC to improve the control and characterization of superconducting hardware~\cite{LeoTalk,QickToolsRepo}. The team is also developing QICK for different qubit platforms such as atomic qubits, spin qubits, and color centers. Preliminary work used the standard QICK firmware~\cite{QICKworkshop,Mounce2023MM}, but the QICK team has since begun creating firmware tailored for these new platforms.

Longer term work on the QICK includes integrating a redesigned timed-processor that will improve the QICK's speed and modularity. This new timed-processor will allow the QICK to scale up to multi-board systems.

\section{Acknowledgements}
This manuscript has been authored by Fermi Research Alliance, LLC under Contract No. DE-AC02- 07CH11359 with the U.S. Department of Energy, Office of Science, Office of High Energy Physics, with support from its QuantISED program and from National Quantum Information Science Research Centers, Quantum Science Center. This work was funded in part by EPiQC, an NSF Expedition in Computing, under grant CCF-1730449. 
The contribution of CZ and MH was supported by the U.S. Department of Energy, Office of Science, National Quantum Information Science Research Centers, Co-design Center for  Quantum Advantage (C2QA) under contract number DE-SC0012704. The work of JM was supported by the NSF Quantum Leap Challenge Institute for Robust Quantum Simulation 2120757 and ARO MURI W911NF-15-1-0397. JM also acknowledges support from NSF
GRFP DGE-2039656.
The contributions of SL, CM, and DIS were supported by the Air Force Office of Scientific Research (AFOSR) Multidisciplinary Research Program of the University Research Initiative (MURI) Grant No. W911NF2010177. CM also acknowledges support from NSF GRFP Grant No. DGE-2146755.
The contributions of CD, HZ, and DIS were supported by the Army Research Office under Grant No. W911NF1910016, and was partially supported by the University of Chicago Materials Research Science and Engineering Center, which is funded by the National Science Foundation under award number DMR-1420709. Devices were fabricated in the Pritzker Nanofabrication Facility at the University of Chicago, which receives support from Soft and Hybrid Nanotechnology Experimental (SHyNE) Resource (NSF ECCS-1542205), a node of the National Science Foundation’s National Nanotechnology Coordinated Infrastructure.
The authors thank National Quantum Information Science Research Centers Q-NEXT, SQMS under contract number DE-AC02-07CH11359, and C2QA members who participated in discussions.

\bibliography{bibliography.bib}

\end{document}